\documentclass[runningheads]{llncs}
\usepackage[T1]{fontenc}
\usepackage{makecell}
\usepackage{url} 
\usepackage{xurl} 
\usepackage{graphicx}
\usepackage{hyperref}
\begin{document}
\title{Detecting Quishing Attacks with Machine Learning Techniques Through QR Code Analysis\thanks{Supported by the Maroun Semaan Faculty of Engineering (MSFEA) at the American University of Beirut (AUB)}}
\titlerunning{Detecting Quishing Attacks with ML Techniques}
%
\author{Fouad Trad\orcidID{0000-0003-2241-8195}\inst{1,2} \and
Ali Chehab\orcidID{0000-0002-1939-2740}\inst{1}}

\authorrunning{F. Trad and A. Chehab}

\institute{
Electrical and Computer Engineering Department, American University of Beirut, Beirut, Lebanon \\
\email{fat10@mail.aub.edu, chehab@aub.edu.lb}
\and
Electrical and Computer Engineering Department, Lebanese American University, Byblos, Lebanon\\
\email{fouad.trad@lau.edu.lb}
} 
\maketitle              
\begin{abstract}
The rise of QR code-based phishing (“Quishing”) poses a growing cybersecurity threat, as attackers increasingly exploit QR codes to bypass traditional phishing defenses. Existing detection methods predominantly focus on URL analysis, which requires the extraction of the QR code payload, and may inadvertently expose users to malicious content. Moreover, QR codes can encode various types of data beyond URLs, such as Wi-Fi credentials and payment information, making URL-based detection insufficient for broader security concerns. To address these gaps, we propose the first framework for quishing detection that directly analyzes QR code structure and pixel patterns without extracting the embedded content. We generated a dataset of phishing and benign QR codes and we used it to train and evaluate multiple machine learning models, including Logistic Regression, Decision Trees, Random Forest, Naïve Bayes, LightGBM, and XGBoost. Our best-performing model (XGBoost) achieves an AUC of 0.9106, demonstrating the feasibility of QR-centric detection. Through feature importance analysis, we identify key visual patterns correlated with phishing labels and refine our feature set by removing non-informative pixels, improving performance to an AUC of 0.9133 with a reduced feature space. Our findings reveal that the structural features of QR code correlate strongly with phishing risk. This work establishes a foundation for quishing mitigation and highlights the potential of direct QR analysis as a critical layer in modern phishing defenses.

\end{abstract}

\section{Introduction}
QR codes have become an integral part of modern digital interactions, facilitating seamless access to websites, payments, authentication systems, and other online services. However, their widespread adoption has also given rise to security threats, particularly in the form of QR code-based phishing attacks, commonly referred to as Quishing \cite{sharevski2022phishing}. In quishing attacks, cyber criminals exploit QR codes to deceive users into scanning them, often leading to credential theft, malware downloads, or financial fraud \cite{vidas2013qrishing}. Unlike traditional phishing attacks that rely on deceptive emails or messages with visible URLs, quishing leverages the opaque nature of QR codes, making it difficult for users to assess their legitimacy before scanning \cite{yong2019survey}.

While URLs are the most common payload in quishing attacks, QR codes can encode a variety of data types beyond web links, broadening the attack surface. They can be used to store Wi-Fi credentials, trigger app deep links, initiate cryptocurrency transactions, add contact details, share geolocation data, send SMS messages, schedule calendar events, or even display plaintext phishing messages. This versatility allows attackers to craft social engineering tactics that do not rely solely on malicious URLs, further complicating detection efforts.

Existing phishing detection methods focus primarily on analyzing URLs and website content, often requiring the resolution of the QR code payload. This approach presents a critical security risk, as accessing the embedded content could expose users or automated security systems to malicious actions before a threat is identified. In addition, obfuscation techniques such as URL shorteners, redirections, and encoded payloads further complicate the detection \cite{le2018using}. Moreover, these techniques are limited to QR codes that embed URLs and do not generalize to QR codes containing other types of data. These challenges highlight the need for an alternative approach that can assess QR code security without resolving its payload.

In this study, we propose, to the best of our knowledge, the first machine learning framework for quishing detection that directly analyzes QR code structure and pixel patterns without relying on URL extraction. Instead of treating QR codes solely as carriers of encoded text, our method leverages their visual and structural properties to distinguish between benign and malicious codes. As such, we created a dataset of QR codes labeled as phishing or benign, and evaluated multiple machine learning models, including Logistic Regression, Decision Trees, Random Forest, Naïve Bayes, LightGBM, and XGBoost. Our results demonstrate that QR-centric detection is not only feasible but also highly effective, with our best-performing model (XGBoost) achieving an AUC of 0.9106.

Furthermore, we conducted a feature importance analysis to identify key visual patterns correlated with phishing labels. By refining our feature set and excluding non-informative pixels, we further improved detection performance to an AUC of 0.9133.

This work establishes a foundation for pre-scan quishing mitigation, demonstrating the viability of direct QR code analysis as a proactive cybersecurity measure. By eliminating the need to extract or resolve QR payloads, whether they contain URLs or other data types, our approach strengthens phishing defense strategies and improves security in a world where QR codes are widely used.

In summary, the main contributions of this paper are as follows:
\begin{itemize}
    \item Proposing a machine learning framework for detecting QR code-based phishing attacks (\textit{quishing}) by directly analyzing QR code structure and pixel patterns without extracting or resolving their embedded content.
    \item Creating a dataset of QR codes (available on \href{https://github.com/fouadtrad/Detecting-Quishing-Attacks-with-Machine-Learning-Techniques-Through-QR-Code-Analysis}{GitHub}) labeled as phishing or benign, enabling the evaluation of QR-centric detection models.
    \item Evaluating multiple machine learning models, including Logistic Regression, Decision Trees, Random Forest, Naïve Bayes, LightGBM, and XGBoost, to assess their performance in quishing detection.
    \item Conducting feature importance analysis on raw QR code pixels to identify critical visual patterns correlated with phishing labels and reducing the feature set to improve performance.
\end{itemize}

The structure of this paper is as follows: Section \ref{relatedwork} provides a comprehensive review of the existing literature on phishing and quishing detection. Section \ref{expsetup} outlines the experimental setup used to create the Quishing dataset, detailing the machine learning models developed, the evaluation metrics employed, and the rationale behind the chosen methodologies. Section \ref{experiments} presents the experiments conducted, followed by a discussion of the results. Finally, Section \ref{conclusion} concludes the paper with a summary of our findings, contributions to the field, and potential avenues for future research.

\section{Related Work}
\label{relatedwork}
Phishing detection has been a critical area of research in cybersecurity, focusing predominantly on identifying malicious URLs through various analytical and machine learning techniques. Traditional approaches to phishing detection have explored features such as URL lexical characteristics, website content analysis, and the use of third-party services like blacklists and WHOIS information to classify URLs as phishing or legitimate. Multiple studies have used machine learning algorithms to analyze URL features, demonstrating significant success in identifying phishing attempts \cite{opara2024look,ahammad2022phishing,hannousse2021towards}.

Further advances have included the implementation of deep learning models, to analyze the textual features of URLs for improved phishing detection accuracy \cite{aljabri2022phishing,wang2023large,tajaddodianfar2020texception}. These techniques have proven effective in identifying malicious URLs by examining their lexical and host-based features and page content to discern patterns indicative of phishing. Very recently, large language models (LLMs) have been explored for this task and are achieving state-of-the-art performance \cite{trad2024prompt,trad_ensemble_2025,trad2024large}.

Despite these advances in phishing detection, the domain of QR code-based phishing, or "quishing," remains largely unexplored. QR codes present unique challenges, as they can encode diverse types of information, allowing attackers to bypass traditional detection mechanisms by embedding malicious content in various forms. Sharevski et al. were among the first to investigate quishing, which gained traction during the COVID-19 pandemic \cite{sharevski2022phishing}. Using a deceptive sign-up for a COVID-19 digital passport, they found that a majority of the 173 participants (67\%) were willing to use their social media credentials for convenience, despite the risks. The study highlighted the threat of quishing, but focused primarily on raising awareness and proposing educational guidelines rather than developing direct countermeasures to mitigate the risk. A study by Amoah and J.B. \cite{amoah2022qr} represents one of the few attempts to address quishing. In their approach, QR codes are first decoded and converted into URLs, which are then subjected to conventional phishing detection methods. Similarly, Rafsanjani et al. introduced QsecR, an Android application that extracts the URL from the QRCode and analyzes it by extracting some static URL features \cite{rafsanjani2023qsecr}. These approaches, while innovative, introduce inherent risks by requiring URL resolution, potentially exposing users to malicious content, and do not exploit the full potential of direct QR code analysis for phishing detection since they are only limited to URLs. 

Our work addresses this gap by introducing a novel machine learning framework that directly analyzes the visual and structural patterns within QR codes, bypassing the intermediary step of payload extraction. By focusing on QR code analysis, we can identify indicators of phishing risk regardless of the type of data encoded. This method provides a proactive pre-scan defense mechanism that automatically assesses the safety of QR codes before they are scanned, thereby enhancing digital security in an era where QR codes serve a multitude of functions.

\section{Experimental Setup}
\label{expsetup}
\subsection{Dataset}
To our knowledge, no publicly available Quishing dataset currently exists. Therefore, a key contribution of this study is the creation of a Quishing dataset, which will be made publicly available to researchers. While QR codes can encode a wide range of payloads, such as Wi-Fi credentials or app links, our dataset focuses specifically on URL-derived QR codes. This focus is motivated by the availability of verified labels (phishing/benign) for URLs in existing phishing datasets, which allows for reliable benchmarking. In contrast, no equivalent datasets currently exist for non-URL QR payloads (e.g., malicious Wi-Fi configurations), as manual verification would require physical device interaction or proprietary threat intelligence. We emphasize that this study does not empirically validate performance on non-URL payloads. Although our method is content-agnostic in principle, extending it to other payload types requires dedicated datasets and remains an important direction for future work.

The dataset we created is derived from the PhishStorm dataset \cite{marchal2014phishstorm} by selecting 10,000 samples, equally categorized into 5,000 legitimate and 5,000 phishing URLs. We then used the 'qrcode' Python library to generate QR codes corresponding to the URLs.

When generating QR codes, we can specify several parameters including:
\begin{itemize}
    \item \textbf{Version:} An integer from 1 to 40, specifying the number of elements in the QR Code.
    \item \textbf{Error Correction:} A choice among 'high', 'medium', and 'low', indicating the degree of error we can tolerate in the QR Code while still navigating to the correct link without errors.
    \item \textbf{Box Size:} The size of an element in pixels (each cell in a QR Code represents a certain number of pixels).
    \item \textbf{Border:} The size of the borders.
\end{itemize}

After investigating the characteristics of our URL data and noting that not every URL can be encoded using any QR code version, we selected version 13 as our standard, which was able to encode most of the URLs (9,987 samples). This choice yields QR codes of a consistent size (69×69 pixels), which is advantageous for machine learning applications, as it allows for uniform pixel-based feature extraction and prediction. Although all QR codes in our dataset are generated with a fixed version (resulting in a uniform 69×69 structure), the internal arrangement of modules still depends on the encoded payload. In particular, properties such as URL length, character composition, and encoding complexity influence how data is distributed within the QR matrix. As a result, the model does not rely on differences in image size, but instead captures variations in internal structural patterns induced by the payload. Moreover, using a single version simplifies the preprocessing pipeline and enhances the comparability of features across samples. The error correction level was set to 'low', which can tolerate up to 15\% loss while still retrieving the URL correctly. Additionally, the box size was set to 1 (minimizing computational demands) and the border to 0 (eliminating extraneous pixels). 

\begin{figure}[h]
\centering
\includegraphics[width=\textwidth]{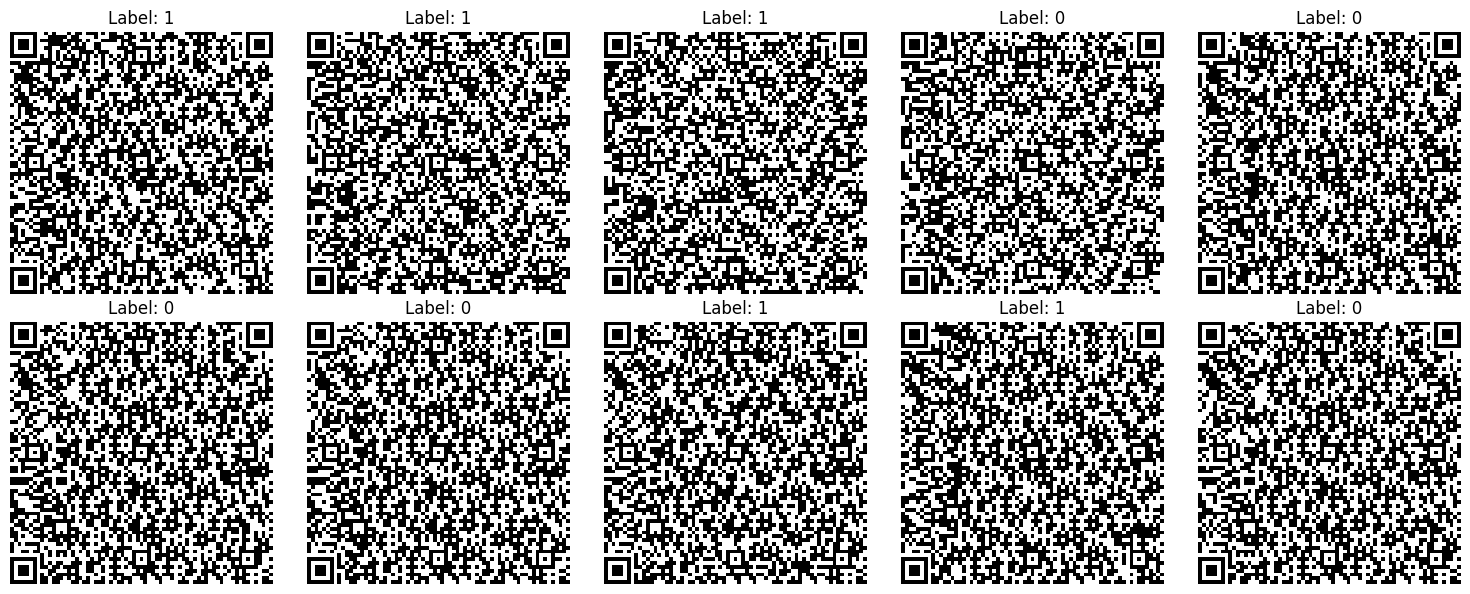}
\caption{Ten samples of the generated QRCodes}
\label{fig:qrcodes}
\end{figure}

We emphasize that the QR codes in our dataset are generated under controlled and uniform parameters (fixed version, error correction level, and rendering settings). This design choice is intentional, as it allows us to isolate the effect of payload-induced structural variations while avoiding confounding factors such as stylistic customization, colorization, or rendering artifacts. While real-world QR codes may exhibit significant variability in appearance, this controlled setup provides a necessary first step to evaluate the feasibility of QR-centric detection. Extending this approach to more diverse and visually varied QR codes remains an important direction for future work. The created dataset can be accessed on GitHub through the following link: 
\url{https://github.com/fouadtrad/Detecting-Quishing-Attacks-with-Machine-Learning-Techniques-Through-QR-Code-Analysis}

Figure \ref{fig:qrcodes} shows 10 samples of the generated QR Codes. The dataset was divided between 80\% training and 20\% testing. We later performed 10-fold cross-validation on the training set to tune the hyperparameters of the used models.

\subsection{Models}
We employ multiple machine learning models, including Logistic Regression, Decision Tree, Naïve Bayes, Random Forest, LightGBM, and XGBoost. Because QR codes adhere to a standardized fixed structure, each QR code image can be reliably flattened into a uniform array, with every pixel treated as an individual feature. This consistency enables effective analysis and comparison across models in a QR code setting. Hyperparameter tuning was performed using a randomized search with 10-fold cross-validation on the training set. The optimal hyperparameters for each model are summarized in Table \ref{tab:hyperparameters} to facilitate reproducibility. While convolutional neural networks (CNNs) and vision transformers (ViTs) are natural choices for image-based tasks, this study focuses on classical machine learning models applied to flattened pixel representations to provide interpretability and establish a baseline for QR-centric detection. This choice allows us to directly analyze feature importance at the pixel level. Exploring deep learning models that capture spatial dependencies is a promising direction for future work.

\begin{table}[h!]
\centering
\caption{Hyperparameters of the chosen models}
\label{tab:hyperparameters}
\begin{tabular}{|l|l|}
\hline
\textbf{Model} & \textbf{Hyperparameters} \\ \hline
Logistic Regression & \texttt{'C': 0.1, 'solver': 'liblinear'} \\ \hline
Decision Tree & \texttt{'max\_depth': 3, 'min\_samples\_leaf': 1} \\ \hline
Random Forest & \texttt{'max\_depth': 20, 'n\_estimators': 100} \\ \hline
LightGBM & \texttt{'learning\_rate': 0.1, 'n\_estimators': 200} \\ \hline
XGBoost & \texttt{'learning\_rate': 0.2, 'n\_estimators': 150} \\ \hline
\end{tabular}
\end{table}


\subsection{Metrics}
The evaluation of our models relied on several classical classification metrics:

\begin{itemize}
    \item \textbf{Accuracy:} This metric measures the overall proportion of correctly classified instances, providing a straightforward assessment of model performance.
    \item \textbf{Precision:} Precision quantifies the proportion of true positive predictions among all positive predictions, indicating the model's ability to minimize false positives.
    \item \textbf{Recall:} Also known as sensitivity, recall measures the proportion of actual positives that are correctly identified by the model, reflecting its capability to detect positive cases.
    \item \textbf{F1-score:} As the harmonic mean of precision and recall, the F1-score offers a balanced evaluation of the model's performance, particularly in scenarios where class distribution is uneven.
    \item \textbf{Area Under the ROC Curve (AUC):} The AUC metric evaluates the model's overall ability to discriminate between classes across different threshold settings, providing a comprehensive measure of classification performance. It is worth noting that the previous metrics are reported based on the default classification threshold of 0.5.
\end{itemize}

\section{Experiments and Results}
\label{experiments}
\subsection{Experiment 1: Training and testing the models}
After identifying the optimal hyperparameters using 10-fold cross-validation, we trained each model on the full training set and evaluated their performance on the testing set. Table \ref{tab:testing_results} presents the computed metrics for each model.

As shown in the results, most models perform well on this task, with the exception of Naïve Bayes, which achieved an AUC of 0.6531, significantly lower than the other models. All other models achieved an AUC above 0.81, demonstrating strong discriminatory ability. Among the tested models, XGBoost, LightGBM, and Random Forest classifiers exhibit the best performance, with AUC scores ranging from 0.8908 to 0.9106.

While AUC provides a threshold-independent measure of model performance, the accuracy and F1-score at the default 0.5 threshold offer additional insights. The LightGBM and XGBoost models achieve the highest accuracy, at 0.8293 and 0.8258, respectively, indicating strong overall classification performance. These models also maintain the highest F1-scores, at 0.8214 and 0.8184, balancing precision and recall effectively. However, it is important to note that classification metrics such as accuracy and F1-score depend on the chosen threshold. Given that AUC is relatively high for these models, adjusting the decision threshold could further optimize the trade-off between precision and recall, particularly in real-world scenarios where minimizing false positives or false negatives is a priority. 

These results suggest that QR code-based phishing detection using direct QR analysis is feasible, with tree-based ensemble models such as LightGBM and XGBoost demonstrating the most promising performance. Further exploration of threshold tuning could refine the detection process based on specific application requirements.

\begin{table}[h!]
\centering
\caption{Performance metrics on the testing set}
\label{tab:testing_results}
\begin{tabular}{|l|c|c|c|c|c|}
\hline
\textbf{Model} & \textbf{Accuracy} & \textbf{Precision} & \textbf{Recall} & \textbf{F1-Score} & \textbf{AUC} \\ \hline
Logistic Regression & 0.7983 & 0.8129 & 0.7621 & 0.7867 & 0.8737 \\ \hline
Decision Tree Classifier & 0.7578 & 0.7358 & 0.7856 & 0.7599 & 0.8138 \\ \hline
Random Forest Classifier & 0.7993 & 0.8570 & 0.7067 & 0.7746 & 0.8908 \\ \hline
Gaussian NB & 0.6376 & 0.8680 & 0.3036 & 0.4498 & 0.6531 \\ \hline
XGBoost Classifier & 0.8258 & 0.8332 & 0.8041 & 0.8184 & 0.9083 \\ \hline
LGBM Classifier & 0.8293 & 0.8394 & 0.8041 & 0.8214 & 0.9106 \\ \hline
\end{tabular}
\end{table}

\subsection{Experiment 2: Deriving Feature Importance}
\begin{figure}[h]
\centering
\includegraphics[width=\textwidth]{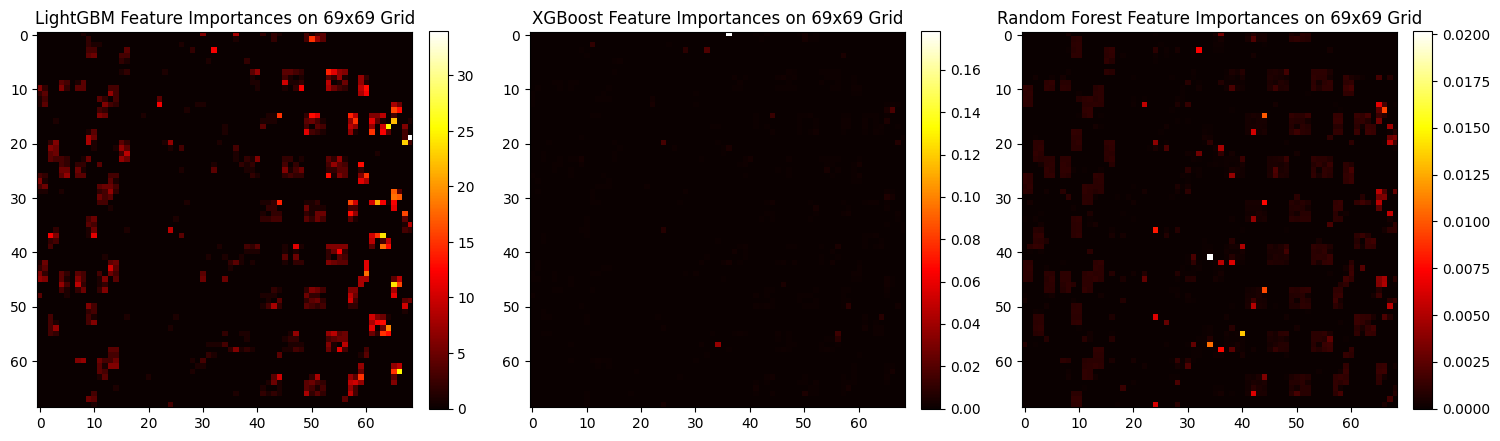}
\caption{Feature Importance of the top 3 models}
\label{fig:feature_importance_top3}
\end{figure}

Based on the results of Experiment 1, we select the top three performing models (Random Forest, LightGBM, and XGBoost) and analyze their feature importance. To visualize the impact of individual pixels on the classification decision, we plot the feature importance values on a 69×69 grid, corresponding to the QR code size, as shown in Figure \ref{fig:feature_importance_top3}. It is important to clarify that the observed feature importance patterns do not imply that QR codes visually encode malicious intent. Rather, the models capture statistical regularities in how different types of payloads (e.g., phishing versus benign URLs) are translated into QR structures through the encoding process.

The results reveal that a significant portion of the QR code remains unused in the prediction process. Large black regions in the feature importance maps indicate that the majority of pixels contribute little to no information for distinguishing phishing from benign QR codes. This suggests that quishing detection is primarily influenced by specific regions of the QR code rather than its full structure.

\begin{figure}[h]
\centering
\includegraphics[width=\textwidth]{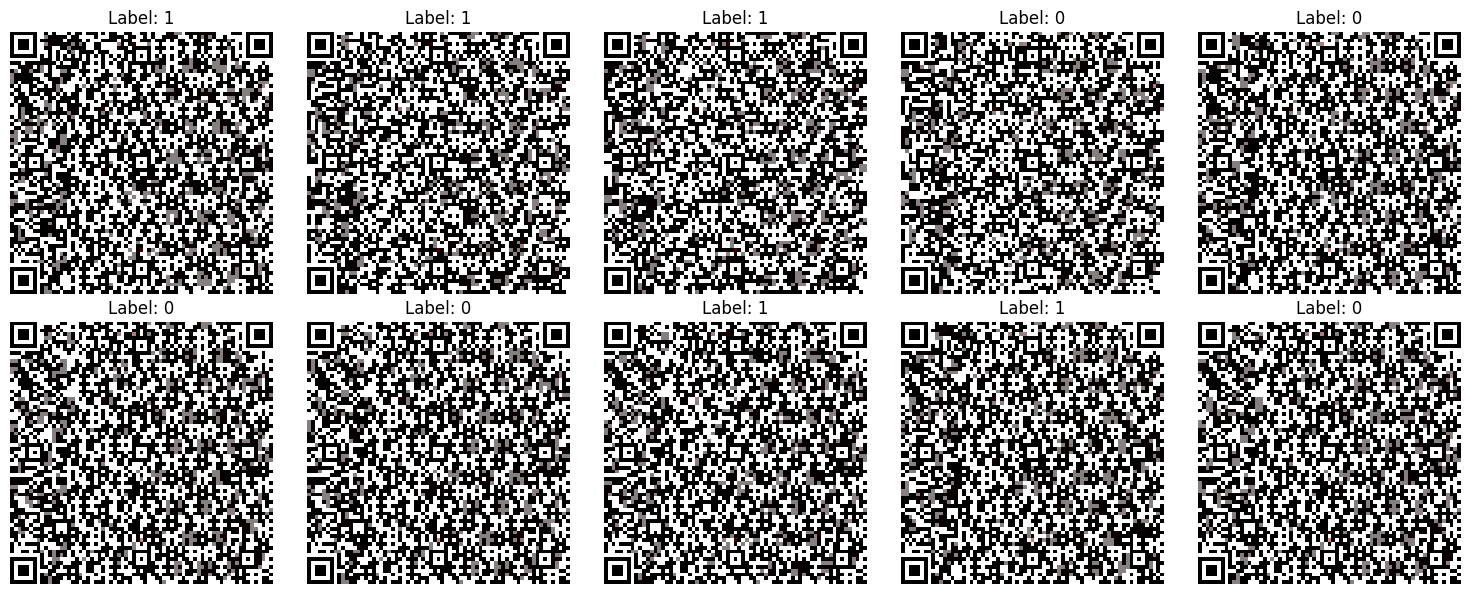}
\caption{Features taken into account when using XGBoost}
\label{fig:features_xgboost_included}
\end{figure}

\begin{figure}[h]
\centering
\includegraphics[width=\textwidth]{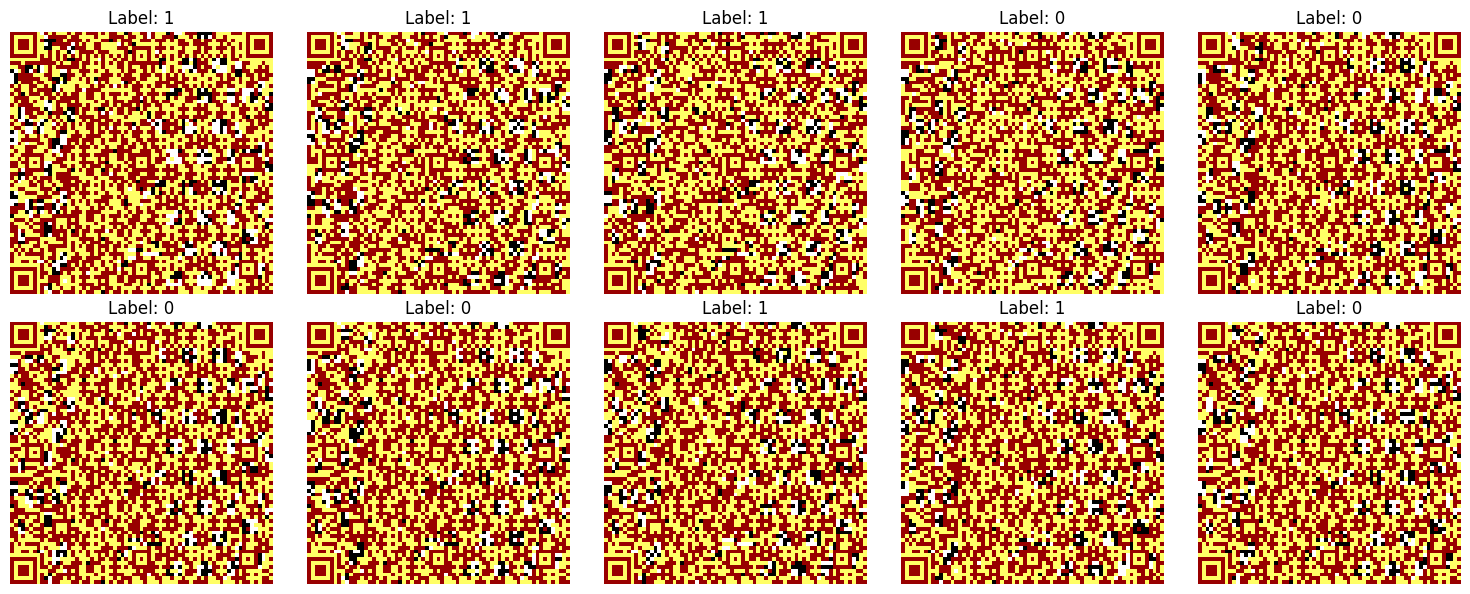}
\caption{Features not taken into account when using XGBoost}
\label{fig:features_xgboost_excluded}
\end{figure}

To further illustrate this observation, we provide two additional visualizations. Figure \ref{fig:features_xgboost_included} highlights the pixels considered important by the XGBoost model, shown in gray, while Figure \ref{fig:features_xgboost_excluded} highlights the excluded pixels, shown in yellow. These figures reinforce the finding that only a subset of pixels is relevant to the classification decision.

\begin{figure}[h]
\centering
\includegraphics[width=\textwidth]{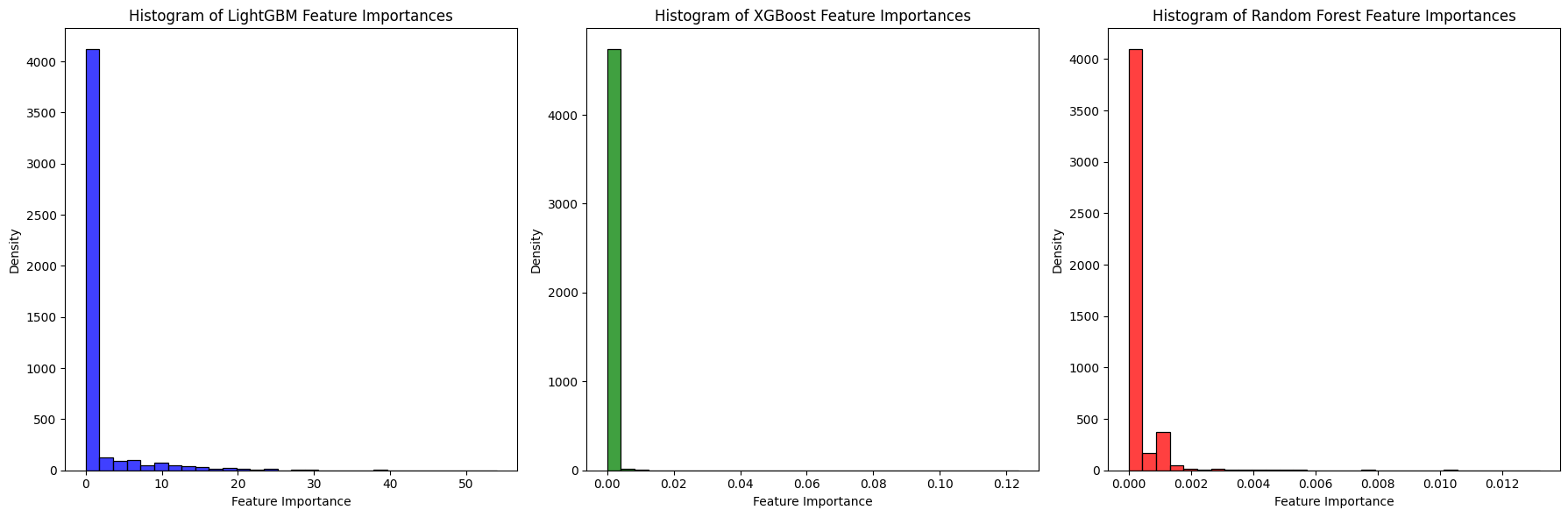}
\caption{Feature Importance distribution}
\label{fig:feature_importance_distribution}
\end{figure}

Additionally, Figure \ref{fig:feature_importance_distribution} presents the distribution of feature importance values for each of the three models. The distribution confirms that the majority of pixels have an importance value of around zero across all models, while only a small fraction contributes meaningfully to the decision-making process. This finding suggests that refining the feature space by focusing on the most relevant pixels could further enhance the model's efficiency and performance.

\subsection{Experiment 3: Feature Selection}
In this experiment, we performed feature selection based on the most important features identified by each of the three best-performing models: Random Forest, LightGBM, and XGBoost. We then retrained all models using only these selected features and compared their performance to their original versions without feature selection. The results, presented in Table \ref{tab:auc_comparison}, show that applying feature selection consistently improves or maintains model performance. This suggests that many pixels in the original QR code images are non-informative and do not contribute to phishing detection.

Among the tested models, Decision Tree is the only one that exhibits no significant change in performance, which is expected given its inherent ability to focus on the most relevant features. The highest performance is achieved with LightGBM, reaching an AUC of 0.9133, further demonstrating the benefits of using a reduced but more informative feature set. 

A particularly notable improvement is observed in Naïve Bayes, where AUC increases from 0.6531 (without feature selection) to as high as 0.8010 when using LightGBM-based feature selection. This suggests that Naïve Bayes, which relies on strong independence assumptions, benefits significantly from a reduced feature set that removes irrelevant or noisy pixels. By limiting the input to only the most important features, the model is less affected by non-informative data, leading to better performance.

Additionally, the impact of different feature selection methods varies slightly across models. For instance, LightGBM achieves its highest AUC (0.9133) when feature selection is based on Random Forest, indicating that the features deemed important by Random Forest align well with LightGBM’s gradient boosting approach. This may be due to Random Forest’s ability to capture a diverse set of relevant features, which LightGBM can then refine further through its boosting mechanism. Similarly, XGBoost shows only slight variations across feature selection methods while consistently maintaining strong performance, reinforcing its robustness to different feature subsets. The relatively small differences in performance across feature selection strategies suggest that all three models are capturing meaningful predictors of phishing risk, further validating the effectiveness of QR code-based feature selection for this task.

\begin{table}[h!]
\centering
\caption{AUC Comparison for Each Model With and Without Feature Selection (FS)}
\label{tab:auc_comparison}
\begin{tabular}{|l|c|c|c|c|}
\hline
\textbf{Model} & \textbf{\makecell{AUC \\ (FS Based on \\ XGBoost)}} & \textbf{\makecell{AUC \\ (FS Based on \\ LightGBM)}} & \textbf{\makecell{AUC \\ (FS Based on \\ Random Forest)}} & \textbf{\makecell{AUC \\ (No FS)}} \\ \hline
Logistic Regression  & 0.8725 & 0.8720 & \textbf{0.8738} & 0.8737 \\ \hline
Decision Tree        & \textbf{0.8138} & \textbf{0.8138} & \textbf{0.8138} & \textbf{0.8138} \\ \hline
Random Forest        & \textbf{0.8984} & 0.8966 & 0.8923 & 0.8908 \\ \hline
Naïve Bayes          & 0.7934 & \textbf{0.8010} & 0.7540 & 0.6531 \\ \hline
XGBoost              & 0.9083 & \textbf{0.9104} & 0.9083 & 0.9083 \\ \hline
LightGBM             & 0.9121 & 0.9106 & \textbf{0.9133} & 0.9106 \\ \hline
\end{tabular}
\end{table}

\section{Conclusion and Future Work}
\label{conclusion}
In conclusion, This study introduced a novel approach to QR code-based phishing (quishing) detection by directly analyzing QR code structure and pixel patterns without relying on URL extraction. Through machine learning models, trained on a newly created dataset, we demonstrate that QR-centric detection is both feasible and effective. Feature selection further improves or maintains model performance by identifying the most informative regions of the QR code while filtering out non-essential pixels. 

While this work establishes an important foundation, there are several avenues for future research. One key direction is to extend the analysis beyond URL-encoded QR codes. Although URLs are the most common vector for quishing attacks, QR codes can also encode other forms of malicious payloads. Developing datasets that encompass these variations will enable a broader assessment of QR code phishing risks and allow models to generalize to a wider range of attack strategies. 

Another promising direction is the incorporation of deep learning techniques, particularly convolutional neural networks and vision transformers, which could further enhance performance by capturing complex spatial patterns within QR codes. While shallow models performed well in this study, deep learning methods may better exploit subtle pixel-based variations indicative of phishing attempts. 

Additionally, while this study operates under controlled conditions, practical implementation requires evaluating models on QR codes captured in varying environments, including different lighting conditions, distortions, and physical printouts. The robustness of QR-based phishing detection systems against adversarial attacks, such as perturbations that deceive machine learning models, is another critical research area. 

From an adversarial perspective, attackers may attempt to manipulate QR code generation parameters (e.g., version, error correction level, or visual styling) to alter structural patterns while preserving functionality. Developing detection methods that are robust to such adversarial variations is an important direction for future research.

Finally, integrating QR code phishing detection into real-world applications is an essential next step. This could involve developing mobile applications, browser extensions, or security software that can assess QR codes before they are scanned, providing users with real-time risk assessments. 

By addressing these challenges, future research can further strengthen defenses against QR code phishing, ensuring that detection methods remain effective against evolving attack techniques. This study provides the groundwork for such advancements, demonstrating that direct QR analysis is a viable and valuable addition to modern phishing mitigation strategies.

%
%
%
\bibliographystyle{splncs04}
\bibliography{mybibliography}

\end{document}